\documentclass[%
reprint,
superscriptaddress,
longbibliography,
amsmath,amssymb,
aps,
prb,
floatfix,
]{revtex4-2}

\usepackage{physics}
\usepackage{graphicx}    
\usepackage{dcolumn}     
\usepackage{bm}          
\usepackage{mathbbol}    
\usepackage{dutchcal}    
\usepackage[dvipsnames]{xcolor}
\usepackage{hyperref}
    \hypersetup
    {   
        unicode=true,           
        pdftoolbar=true,        
        pdfmenubar=true,        
        pdffitwindow=false,     
        pdfstartview={FitH},    
        pdfauthor={Gabriele Bellomia, Adriano Amaricci, Massimo Capone},
        pdftitle={Local classical correlations between physical electrons in Hubbard systems},
        pdfkeywords={strong correlations, nonfreeness, natural orbitals, entanglement measures, quantum information theory, dynamical mean-field theory, quantum embedding, open quantum system, magnetic ordering, metal-insulator transition},
        pdfnewwindow=true,      
        colorlinks=true,        
        linkcolor=MidnightBlue,        
        citecolor=Maroon,        
        filecolor=Brown,        
        urlcolor=Maroon          
    }

\renewcommand\labelenumi{(\roman{enumi})}
\renewcommand\theenumi\labelenumi

\newcommand\eg{\mbox{\it e.g.\ }}
\newcommand\ie{\mbox{\it i.e.\ }}

\newcommand\up{\uparrow}
\newcommand\dw{\downarrow}

\begin{document}

\title{Local classical correlations between physical electrons in Hubbard systems}

\author{Gabriele Bellomia}
    \email{gabriele.bellomia@sissa.it}
    \affiliation{{\sc SISSA}, Scuola Internazionale Superiore di Studi Avanzati, Trieste, Italy}

\author{Adriano Amaricci}
    \affiliation{{\sc CNR--IOM}, Istituto Officina dei Materiali, Consiglio Nazionale delle Ricerche, Trieste, Italy}

\author{Massimo Capone}
    \affiliation{{\sc SISSA}, Scuola Internazionale Superiore di Studi Avanzati, Trieste, Italy}
    \affiliation{{\sc CNR--IOM}, Istituto Officina dei Materiali, Consiglio Nazionale delle Ricerche, Trieste, Italy}

\begin{abstract}
We demonstrate that the local {\it nonfreeness}, an unbiased measure of correlation between electrons at a single lattice site, can be computed as the mutual information between local natural spin orbitals.
This leads us to prove a general result: local electron correlations in Hubbard-type models that conserve the orbital- and spin-resolved electron number are fully classical, since the local reduced density matrix is separable in the natural basis and no quantum correlations beyond entanglement are present. Finally, we compare different theoretical descriptions of magnetic and nonmagnetic states, showing that local classical correlations are drastically influenced by nonlocal processes.
These results confirm the relation between local classical correlations within an open system and nonlocal entanglement and they provide a clear path for the study of the relationship between traditional quantum resources and the nonfreeness in terms of experimentally accessible quantities.
\end{abstract}

\maketitle


\section*{Introduction}
The concept of strongly correlated electrons is widely used to identify a broad class of systems that display a variety of phases and regimes arising from the electron-electron interactions. Given their rich and tunable phenomenology, strongly correlated electron
systems have been identified as promising platforms for quantum technologies, which have
assumed a central role in the second quantum revolution.
However, a quantitative measure of strong particle correlations that can differentiate the quantum and classical contributions is still elusive, and different estimates can be used in different frameworks. Some of these estimates are not uniquely defined as one can use different reference states or orbital bases.
For instance, in variational methods like density-functional theory or quantum Monte Carlo, the \emph{correlation energy} (energy difference with respect
to a Slater determinant) is a natural candidate \cite{Wigner_corr_energy,ParrYang,VMC_corr_energy,strong_corr_puzzle},
whereas configuration interaction schemes naturally ascribe correlations to the number of necessary Slater determinants~\cite{ASCIdmft,Arrigoni_CI,Ganoe_Faraday_corr_defs}. 
In diagrammatic methods one can define the functional distance between the self-energy and/or two-particle vertex and their uncorrelated
mean-field (Hartree-Fock) counterparts \cite{AA_2015,Millis_corr_self-energy,Gull_corr-self-energy,Chalupa_Vertex_Divergence,Krien_corr_self-energy}. 
Finally tensor network methods rely on the notion of bond dimension~\cite{TenNet_RMP,Carisch2023}. The latter is deeply rooted in the framework of quantum information theory in Fock space, which we will refer to as \emph{orbital information theory} in the following, as it is concerned with correlations and entanglement between (groups of) single-particle
states (orbitals), rather than correlations between the constituent physical particles~\cite{EntanglementRMP,Laflorencie_Review,Vedral2001,Banhuls2007,Benatti2020,Schilling_QuantumScience}.

Building on ideas recently introduced in different communities~\cite{BellomiaPhD,Faraday_Natural_Orbitals}, here 
we propose an unbiased measure of the correlations contained in a single
electronic orbital (or intraorbital correlations) as the mutual information between its two \emph{natural} spin orbitals 
that  define the single bipartition of the single-orbital
density matrix into eigenstates of the single-electron density matrix (up and down, when the spin is quantized along the $z$ direction).
The recipe is readily generalized including $N$ local orbitals by considering
a $2N$-partition into the $2N$ natural spin orbitals. Here we 
focus on the $N=1$ case for its simplicity and relevance for the study
of many systems in which the one-electron density matrix does not mix the
atomic orbitals, including orbital-selective phases and interaction-resilient metals
\cite{LdM2005_orb-selective,Medici2011PRB,Janus2011,Medici2014PRL,Isidori2019,DelRe2018_selective,Richaud2021PRB,Richaud2022,Tusi2022NatPhys}.

We first prove that, under completely general circumstances, for a state that conserves the total 
(orbital-resolved\,\footnote{For a multiband state we require 
the density and magnetization in each band to be conserved independently})
density and magnetization, all local intraorbital correlations are 
\emph{effectively classical}, as neither entanglement~\cite{Vedral_RMP,NielsenChuang,Peres96,Horodecki97,HHHH_RMP,EntanglementDetection_review,Plenio_EntanglementMeasures} 
nor quantum discord~\cite{Discord_def,Discord_2qbits,Discord_conditions,Discord&Classical_unified,Modi_RMP,Discord_RMP[qfromclassical],Discord_REE,Global_Discord_Multipartite,Multipartite_Discord} 
between natural spin orbitals is contained in the local reduced density matrix (RDM).
Then we prove that our measure of intraorbital classical correlations vanishes for any
Hartree-Fock state and it coincides with the single-orbital \emph{nonfreeness}, 
a well-known quantitative measure of correlations between physical electrons
\cite{Gottlieb2005,Gottlieb2007,Held_nonfreness,Vollhardt_nonfreeness}.
This leads us to a rigorous and unbiased measure of local classical correlations which is based on simple quantities that are experimentally accessible in quantum simulators. 

We follow this direction studying the half-filled Hubbard model on the square and honeycomb lattices. Comparing local embedding methods with numerically exact approaches, we show that the local classical nonfreeness captures the physics of strong correlations in an unbiased way and, remarkably, it is clearly
sensitive to nonlocal quantum fluctuations and correlations, as expected from the quantum information theory of open systems \cite{Discord_RMP[qfromclassical],Artiaco_PRL2025}.

\section*{Local spin-orbital correlations}

Let us consider a lattice model of itinerant fermions in which the 
orbital-resolved density and magnetization are conserved, as realized \eg in the single-band 
Hubbard model~\cite{Hubbard1963}, its SU($N$) versions \cite{ultracold_SU(N)}, and the density-density version of the Kanamori 
Hamiltonian \cite{Kanamori}, where the rotational symmetry is partially broken.
In light of the abelian nature of such symmetries, we can write the RDM describing a single lattice site $\rho_\mathrm{loc}$ so 
as to commute with the {local}, \emph{orbital-resolved}, density and magnetization. The explicit derivation can be found in the appendix.
Considering an arbitrary number of orbitals, labeled as $i=1,2,\dots,N$, the orbital-resolved conservation of
spin densities $\langle n_{i,\sigma} \rangle$ ensures a diagonal form also for the full $N$-orbital local density matrix $\rho_\mathrm{loc}$, 
involving expectation values of $2N$-particle operators, of the form $\langle n_{1,\sigma} n_{2,\sigma'} \cdots n_{N,\sigma''} \rangle$.
It follows that the local RDM can be written as a mixture of products of pure states
\begin{align}
    \rho_\mathrm{loc} &= \sum_{p} p \bigl( \lvert \psi_{1,\up}^{p} \rangle \langle \psi_{1,\up}^{p} \rvert \otimes \lvert \psi_{1,\dw}^{p} \rangle \langle \psi_{1,\dw}^{p} \rvert \otimes \cdots \nonumber \\
    &\qquad\quad \cdots \otimes \lvert \psi_{N,\up}^{p} \rangle \langle \psi_{N,\up}^{p} \lvert \,\otimes\, \lvert \psi_{N,\dw}^{p} \rangle \langle \psi_{N,\dw}^{p} \rvert \bigr),
    \label{eq:rho_loc}
\end{align}
which are separable with respect to the multipartition in the $2N$ spin orbitals 
\cite{NielsenChuang,Peres96,Horodecki97,HHHH_RMP,Cirac_multipartite_separable}
and do not contain any global multipartite quantum discord 
\cite{Discord_def,Discord_2qbits,Discord_conditions,Discord&Classical_unified,Modi_RMP,Discord_RMP[qfromclassical],Discord_REE,Global_Discord_Multipartite,Multipartite_Discord}.
Hence all local spin-orbital correlations are determined by the classical probability $\{p\}$. Remarkably, they can be accurately measured by the multipartite mutual information between the $2N$ separable subsystems \cite{Faraday_Natural_Orbitals}.

To facilitate the contact with experiments, we now specialize the result to a single orbital $i$, 
addressing its bipartite spin-orbital correlations in terms
of easy to compute and measurable quantities. For SU($N$) systems, where all flavors are inherently equivalent, this would give a fairly complete picture of local classical correlations, while for more complex systems with nonequivalent orbitals (as \eg orbital-selective phases) one would
evaluate the correlation of physically distinct orbitals separately.
As a direct consequence of Eq.\,(\ref{eq:rho_loc}), the RDM describing
a single local orbital $i$ takes the following diagonal form~\cite{XiDai2013,WalshLetter2019,WalshPRXQuantum2020,WalshPNAS2021,Held_2RDMfrom2GF}:
\begin{equation}
    \rho_i = 
    \sum_{\ell=\varnothing,\up,\dw,\up\dw}p_\ell\ketbra{\ell},
     \label{eq:zrho}
\end{equation}
where the $p_{\ell}$ coefficients define a classical probability distribution for the four
local states $\ket{\varnothing}=\ket{0}\otimes\ket{0}$, $\ket{\up}=\ket{1}\otimes\ket{0}$,
$\ket{\dw}=\ket{0}\otimes\ket{1}$ and $\ket{\up\dw}=\ket{1}\otimes\ket{1}$, 
expressed in the natural occupation basis $\ket{n_\up}\otimes\ket{n_\dw}$ \footnote{The natural spin orbitals are the eigenstates of any spin projection, \eg the $S_z$ basis, in the Hubbard model.}.
We can explicitly express $\rho_i$ in terms of the local doping fraction, magnetization
and double occupancy, respectively $\delta_i = 1 - \expval{n_{i,\up}}-\expval{n_{i,\dw}}$,
$\mathcal{m}_i = \expval{n_{i,\up}}-\expval{n_{i,\dw}}$ and  $\mathcal{D}_i = \expval{n_{i,\up} n_{i,\dw}}$, evaluated on the $i$-th orbital: 
\begin{equation}
\label{eq:local_rho_matrix}     
\rho_i =
\begin{pmatrix}
\delta_i\!+\!\mathcal{D}_i & 0 & 0 & 0 \\[6pt]
0 &
\displaystyle \frac{1\!-\!\delta_i\!+\!\mathcal{m}_i}{2}\!-\!\mathcal{D}_i &
0 & 0 \\[10pt]
0 & 0 &
\displaystyle \frac{1\!-\!\delta_i\!-\!\mathcal{m}_i}{2}\!-\!\mathcal{D}_i &
0 \\[10pt]
0 & 0 & 0 & \mathcal{D}_i
\end{pmatrix}.
\end{equation}

Since Eq.\,(\ref{eq:zrho}) describes a classical mixture of product states, 
hence a separable state with respect to the bipartition into the two 
spin orbitals, these latter are not entangled for any value of the parameters~\cite{NielsenChuang,Peres96,Horodecki97,HHHH_RMP}. 
Furthermore, the diagonal structure of Eq.\,(\ref{eq:zrho}) in the natural basis guarantees that all bipartite correlations between the two spin orbitals are only due to the probability distribution $\{p_\ell\}$. Hence, they should be considered classical as there is no (bipartite) quantum discord~\cite{Discord_def,Discord_2qbits,Discord_conditions,Discord&Classical_unified,Modi_RMP,Discord_RMP[qfromclassical],Discord_REE}.

In direct analogy to the multipartite case, local classical correlations contained in a single orbital 
are quantified by the \emph{intraorbital mutual information} 
between the selected local spin orbitals~\cite{BellomiaPhD} (\ie between the $\up$ and $\dw$ single-electron states on the same orbital):
\begin{equation}
    I(\,\up\,:\,\dw\,) ~=~ s(\rho_{i,\up}) + s(\rho_{i,\dw}) - s(\rho_i)\,,
    \label{eq:intracorr}
\end{equation}
where $s(\rho)=-\trace\rho\log(\rho)$ is the von Neumann entropy of a state $\rho$, and $\rho_{i,\up}$ and $\rho_{i,\dw}$ are the
two local spin-orbital density matrices, defined by the partial traces:
\begin{equation}
\begin{aligned}
    \rho_{i,\up} =  \trace_\dw \rho_i = 
                    \begin{pmatrix} 
                    \displaystyle \frac{1\!+\!\delta_i\!-\!\mathcal{m}_i}{2} & 0 \\[8pt] 
                    0 & \displaystyle \frac{1\!-\!\delta_i\!+\!\mathcal{m}_i}{2} 
                    \end{pmatrix}, \\[1mm]
    \rho_{i,\dw} =  \trace_\up \rho_i = 
                    \begin{pmatrix}
                    \displaystyle \frac{1\!+\!\delta_i\!+\!\mathcal{m}_i}{2} & 0 \\[8pt] 
                    0 & \displaystyle \frac{1\!-\!\delta_i\!-\!\mathcal{m}_i}{2} 
                    \end{pmatrix}. \label{eq:spin_rho} 
\end{aligned}
\end{equation}
The mutual information $I(\,\up\,:\,\dw\,)$ gives in fact an upper bound for all possible correlation functions between the two spin orbitals~\cite{Wolf}, namely
\begin{equation}
   I(\,\up\,:\,\dw\,) \geq \frac{\left(\expval{\mathcal{O}_\up \otimes \mathcal{O}_\dw}_{\rho_i} - \expval{\mathcal{O}_\up}_{\rho_{i,\up}}\expval{\mathcal{O}_\dw}_{\rho_{i,\dw}}\right)^2}{2\norm{\mathcal{O}_\up}^2\norm{\mathcal{O}_\dw}^2},
   \label{eq:correlation_inequality}
\end{equation}
for any pair $\mathcal{O}_\up$, $\mathcal{O}_\dw$ of generic
operators acting on the Hilbert spaces of subsystems associated to the two spin 
and with $\|\mathcal{O}_\circ\|$ denoting the Euclidean operator norm of $\mathcal{O}_\circ$ (its largest singular value).

So far, we established that two (or more) local spin orbitals are never entangled
in Hubbard-like models that conserve the orbital-resolved density and magnetization and
we provided a general and accessible recipe to quantify their
classical correlations, in terms of the intraorbital mutual information. 
We end the section by showing that in the Hartree-Fock mean-field theory also these 
intraorbital correlations vanish. This is easily proven 
from the fact that 
the double occupancy is factorized in the product of single-particle expectation values, 
namely 
$\mathcal{D}_i^\mathrm{HF} = \expval{n_{i,\up}}\expval{n_{i,\dw}}
= \frac{1}{4}\left({1-\delta_i+\mathcal{m}_i}\right)\left({1-\delta_i-\mathcal{m}_i}\right)
$, which plugged into  Eq.\,(\ref{eq:local_rho_matrix}) gives 
$\rho_i^\mathrm{HF} = \rho_{i,\up}\otimes\rho_{i,\dw}$ corresponding to a vanishing intraorbital mutual information.
Indeed, from a quantum information geometry point of view~\cite{Vedral_RMP, Schilling_QuantumScience,BellomiaPhD},
the mutual information is defined as the distance between the quantum state $\rho_\mathrm{AB}$ and the closest product state
$\rho_\mathrm{A}\otimes\rho_\mathrm{B}$ for any partition into A and B subsystems. 
This implies that the Hartree-Fock approximation to $\rho_i$ corresponds to the optimal 
product state with respect to the bipartition in local natural orbitals.
In this sense Eq.\,(\ref{eq:intracorr}) provides a notion of statistical distance between 
a given $\rho_i$ and the Hartree-Fock solution.

\section*{Equivalence to local nonfreeness}

A fundamental quantity in the framework of electronic structure theory is the nonfreeness~\cite{Gottlieb2005,Gottlieb2007,Held_nonfreness,Vollhardt_nonfreeness}, 
an unbiased quantitative measure of correlations between electrons, defined 
as the quantum relative entropy~\cite{Vedral_RMP,Schilling_QuantumScience,BellomiaPhD} between a given state $\rho$ and the set of 
free states $\rho^\mathrm{free}$, \ie the Gibbs ensemble associated to the closest Hartree-Fock theory~\footnote{Alternatively, one can define free states as those
for which the Wick theorem holds for computing any $N$-body expectation value. Hence,
the closest free state is the best one-body approximation of the $N$-body state.}: 
\begin{equation}
    \mathfrak{N}(\rho) = \min_{\{\rho^\mathrm{free}\}}{S(\rho\,||\,\rho^\mathrm{free})}.
    \label{eq:nonfree_def}
\end{equation}
Remarkably, the minimization in Eq.\,(\ref{eq:nonfree_def}) can be performed analytically~\cite{Gottlieb2005,Gottlieb2007,Held_nonfreness},
characterizing the unique closest free state as to have the same one-electron
density matrix $\gamma_\rho = \langle c^\dagger_{i,\sigma}c_{j,\sigma'}\rangle_{\!\rho}$ as the
many-electron state $\rho$. The resulting closed expression is given as~\cite{Faraday_Natural_Orbitals}:
\begin{equation}
    \mathfrak{N}(\rho) = s(\gamma_\rho) + s(\mathbb{1}-\gamma_\rho)-s(\rho).
    \label{eq:nonfree_closed}
\end{equation}

Despite its success in quantifying in a reasonable and well-defined way 
correlations between electrons~\cite{Held_nonfreness,Vollhardt_nonfreeness},
the nonfreeness cannot be easily connected to
the quantum information theory of many-body systems. 
The latter, in fact, assumes the subsystems of interest to be distinguishable whereas it is well known that individual quantum particles are in general indistinguishable, especially 
in condensed matter settings. 
For this reason, it has been long believed that the inter-particle correlations measured by
the nonfreeness could not be split in quantum and classical parts as
done, for instance, in the framework of orbital information theory~\cite{Schilling_QuantumScience}. 
However, a rather general
argument was recently proposed 
stating that the 
minimization procedure in Eq.\,(\ref{eq:nonfree_def}) is equivalent to minimize 
the mutual information between spin orbitals, with respect to all possible orbital bases.
The resulting optimal single-electron basis corresponds to the \emph{natural} spin orbitals
of the given system, \ie the eigenvectors of $\gamma_\rho$~\cite{Faraday_Natural_Orbitals}.
While the general proof is rather involved, as it assumes that the
given $\rho$ satisfies the fermionic parity superselection rule \cite{Wick1952,Wick1970,SSR_lecturenotes,Friis2013,Friis2016,Galler_SSR,Bellomia2023,DelMaestro_SSR}
and exploits results from majorization theory \cite{Nielsen_majorization,Marshall_majorization},
we can explicitly show the equivalence of $I(\,\up\,:\,\dw\,)$ [Eq.\,(\ref{eq:intracorr})] and 
$\mathfrak{N}(\rho_i)$ [Eq.\,(\ref{eq:nonfree_closed})],
leveraging the symmetry properties of the single-orbital density matrix $\rho_i$ in Eq.\,(\ref{eq:zrho}).
Given that $s(\rho_i)$ enters both expressions additively, we are left to prove that $s(\rho_{i,\up})+s(\rho_{i,\dw}) = s(\gamma_{\rho_i}) + s(\mathbb{1}-\gamma_{\rho_i})$. 
Given
\begin{equation}
\begin{aligned}
    \gamma_{\rho_i} = 
    \begin{pmatrix}
        \expval{n_{i,\up}} & 0 \\
        0 & \expval{n_{i,\dw}}\\
    \end{pmatrix} &= 
    \begin{pmatrix}
    \displaystyle \frac{1\!-\!\delta_i\!+\!\mathcal{m}_i}{2} & 0 \\[8pt] 
    0 & \displaystyle \frac{1\!-\!\delta_i\!-\!\mathcal{m}_i}{2} 
    \end{pmatrix}, \\[3mm]
    \mathbb{1} - \gamma_{\rho_i} &= 
    \begin{pmatrix} 
    \displaystyle \frac{1\!+\!\delta_i\!-\!\mathcal{m}_i}{2} & 0 \\[8pt] 
    0 & \displaystyle \frac{1\!+\!\delta_i\!+\!\mathcal{m}_i}{2} 
    \end{pmatrix}, \label{eq:1bdm}
\end{aligned}
\end{equation}
a direct comparison of all diagonal entries in Eqs.\,(\ref{eq:spin_rho})
and (\ref{eq:1bdm}) completes the proof. 
Given that $\rho_i$ contains
only classical correlations between the natural orbitals underlying the definition of
$I(\,\up\,:\,\dw\,)$, we conclude that, whenever the orbital-resolved spin density is conserved, 
the local nonfreeness measures only classical correlations. In the more general case, one can
exploit the bipartition into natural spin orbitals to minimize the quantum relative entropy
with respect to separable states or pseudo-classical/pointer states~\cite{Discord_def,Discord_2qbits,Discord_conditions,Discord&Classical_unified,Modi_RMP,Discord_RMP[qfromclassical],Discord_REE},
instead of product/free states, and so find measures of entanglement and quantum discord
between physical electrons in the state $\rho_i$. \\
In other words, the power of the 
$I(\,\up\,:\,\dw\,)\equiv\mathfrak{N}(\rho_i)$
correspondence lies in the definition of an optimal
bipartition into single-electron states and so the access to particle correlations within
the  well-developed framework of orbital quantum information theory
\cite{EntanglementRMP,Laflorencie_Review,Banhuls2007,Vedral2001,Schilling_QuantumScience}.

In the more general case of a local RDM $\rho_\mathrm{loc}$ consisting of $N$ charge and magnetization conserving
orbitals, one can easily generalize the proof, without invoking any superselection rule: the symmetry properties
of $\rho_\mathrm{loc}$ ensure that the single spin-orbital density matrices $\rho_{i,\sigma}$ are all diagonal, 
generalizing Eq.\,(\ref{eq:spin_rho}), and their eigenvalues correspond to the corresponding natural occupations 
$\langle n_{i,\sigma}\rangle$ and $1 - \langle n_{i,\sigma}\rangle$.
Hence, the multipartite mutual information between all spin orbitals 
\cite{Faraday_Natural_Orbitals} matches the full local nonfreeness:
\begin{equation}
    \mathfrak{N}(\rho_\mathrm{loc}) \equiv \sum_{i=1}^N \bigl[ s(\rho_{i,\up}) + s(\rho_{i,\dw}) \bigr] - s(\rho_\mathrm{loc}),
\end{equation}
proving the classical nature of local interparticle correlations also at the multipartite level.

\section*{Mott-Hubbard transition}
\begin{figure}
    \centering
    \includegraphics[width=.85\linewidth]{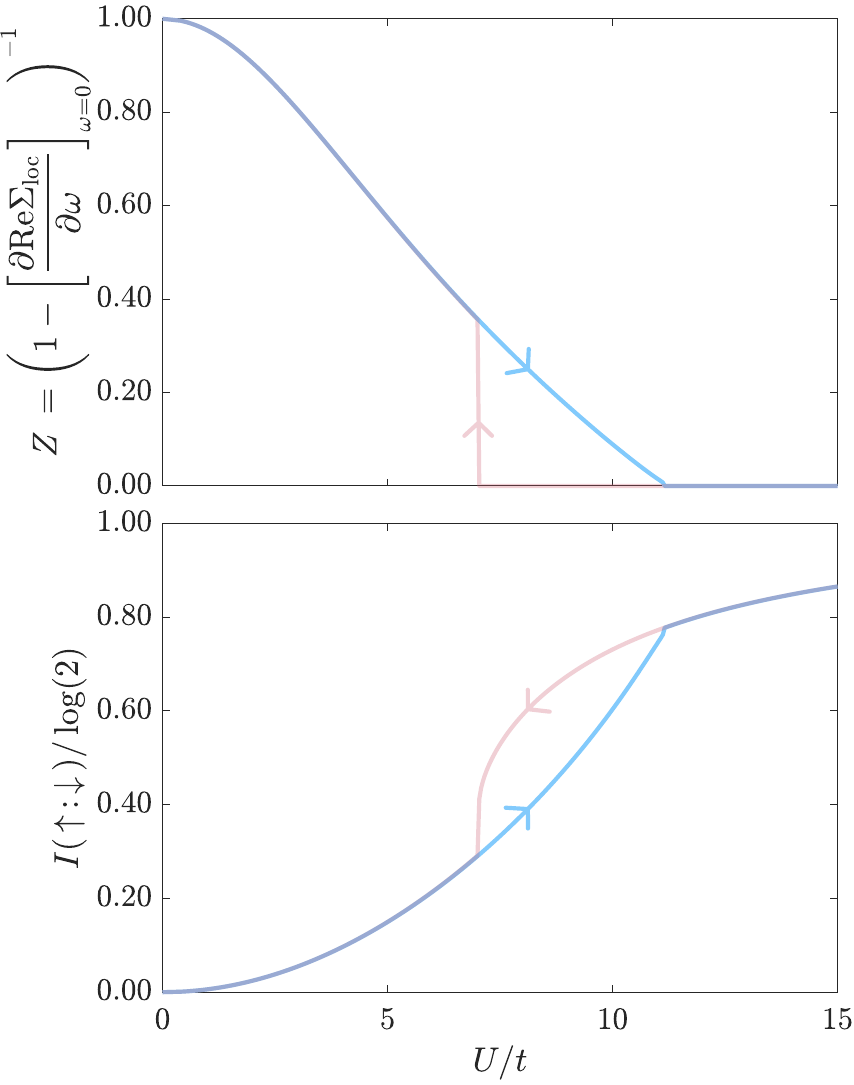}
    \caption{Local quasiparticle weight (top) and local nonfreeness (bottom) across the paramagnetic
    Mott-Hubbard transition on the square lattice, at zero temperature, in the gRISB approximation.
    The arrows depict the increasing-$U$ and decreasing-$U$ solutions.}
    \label{fig:square_mit}
\end{figure}

Having defined a rigorous and general measure of local classical correlations between electrons, Eq.\,(\ref{eq:intracorr}), in the following we showcase its power to provide further insight and understanding about strongly correlated fermions.
We consider the most characteristic fingerprint of electron-electron correlations: the interaction-driven transition between a metal and a Mott insulator in the single-band half-filled Hubbard model~\cite{Mott1949,Mott_RMP,Castellani,Brinkman1970,Vollhardt1984,DMFT_RMP}. 

The Hubbard model on a square lattice has an antiferromagnetic (AFM) ground state for every value of $U$. However, we start from the paramagnetic (PM) solution, in which magnetism is inhibited to single out the intrinsic effects of the local repulsion.
We address this problem within a local quantum embedding, which describes the effect of correlations in term of a local self-energy: the ghost rotational invariant slave bosons (gRISB) method (equivalent to the ghost-Gutzwiller)~\cite{Lanata_EmergentBloch,Lanata_gGut_GF}. 
While the RISB~\cite{RISBoriginal,Gebhard_RISB=Gutzw,RISB_Isidori2009,RISBunified} method only captures the low-energy itinerant properties of correlated electrons, gRISB includes a description of high-energy excitations by introducing extra auxiliary degrees of freedom, and produces
results of accuracy comparable to dynamical mean-field theory (DMFT) \cite{DMFT_RMP} at a much smaller computational cost \cite{Lanata_gRISB_vs_DMFT, Guerci2019,Mejuto_gGut_multiorbital,Giuli_Altermagnet_gRISB,Tagliente2025,TopoGhost2025,AIM1961,Guerci2023}. 
In the present context, we note that gRISB faithfully represents the local RDM \cite{RISBunified,GiuliPhD}, and hence it allows us to compute straightforwardly the intraorbital mutual information in Eq.\,(\ref{eq:intracorr}).

In the upper panel of Fig.\,\ref{fig:square_mit} we show the quasiparticle weight $Z$
as a function of the ratio $U/t$ for the square lattice at zero temperature. 
$Z$ is maximum for $U=0$ and it decreases with increasing $U$ up to a critical value $U_\mathrm{c2}$, where it vanishes continuously, signaling a metal-insulator transition.
Starting from large $U$ and reducing the interaction strength, we recover a metal for $U\leq U_\mathrm{c1}$, where  $Z$ abruptly jumps to a finite value.
The whole scenario is in good agreement with the well established picture from DMFT \cite{DMFT_RMP}.

We now focus on the new insight brought by $I(\,\up\,:\,\dw\,)$. This measure of correlations vanishes in the noninteracting limit and grows with $U$, approaching in the strong-coupling regime ($U\gg t$) the value of $\log(2) =$ 1 bit, \ie the maximum possible value for classical correlations between two qubits. $I(\,\up\,:\,\dw\,)=\log(2)$ is
reached only in the atomic limit $t=0$, where the double occupancy is exactly zero. 
Similarly to $Z$, also the local classical correlations display a hysteresis in the region $U_\mathrm{c1} \leq U \leq U_\mathrm{c2}$ and both solutions are singular at their respective critical interactions.
We stress that for $U > U_{c2}$ $I(\,\up\,:\,\dw\,)$ is close to the maximum value of 1 bit, revealing the strongly-correlated nature of the paramagnetic Mott-Hubbard insulator.
Recent work \cite{Bellomia2023,BellomiaPhD} has proposed a \emph{quasilocal} (short-range nonlocal)
quantum origin to these maximal local classical correlations, in the form of superselected entanglement between nearest-neighboring sites. 

\section*{Antiferromagnetic state}

\begin{figure}
    \centering
    \includegraphics[width=\linewidth]{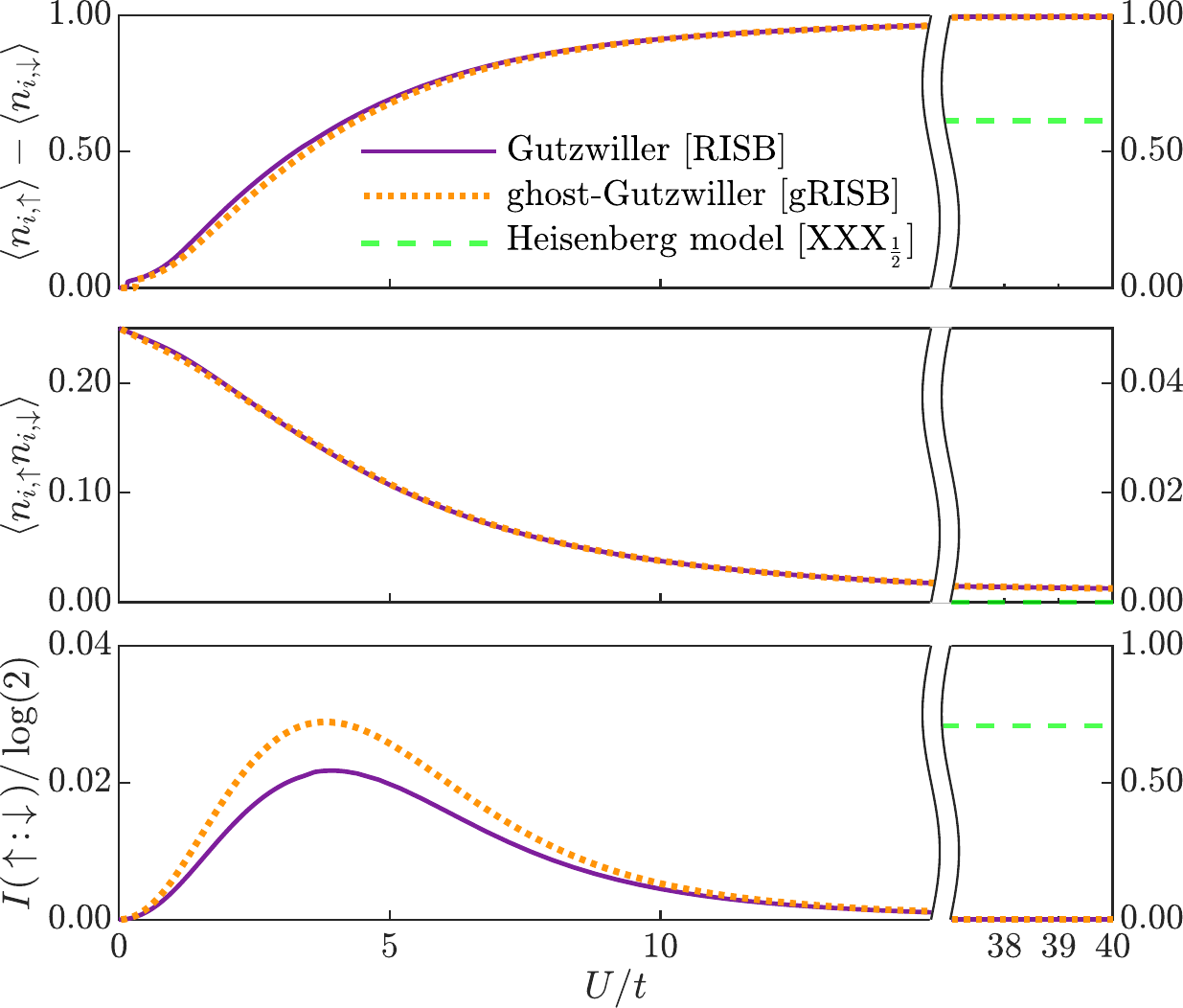}
    \caption{Local magnetization (top), local double occupancy (center) and local nonfreeness
    (bottom) in the ground state of the Hubbard model on the square lattice, comparing the RISB
    approximation and the gRISB approximation ($N_\mathrm{ghost}=2$).
    The dashed green line in the strong-coupling regime marks the exact solution of the isotropic
    Heisenberg model \cite{Sandvik_squareXXX_2002}.}
    \label{fig:square_afm}
\end{figure}

We proceed by studying the AFM ground state on the square lattice.
In Fig.\,\ref{fig:square_afm} we present the local magnetization $\mathcal{m}_i$ (top), the local double occupancy $\mathcal{D}_i$ (center) and our measure of local classical
correlations $I(\,\up\,:\,\dw\,)$ (bottom), comparing results from RISB, gRISB, and numerically exact results for the spin-$\frac{1}{2}$ isotropic Heisenberg model (XXX$_\frac{1}{2}$)~\cite{Sandvik_squareXXX_2002,Sandvik_honeyXXX_2006}, corresponding to the large-$U$ limit of the half-filled Hubbard model.
The value of $I(\,\up\,:\,\dw\,)$ in the XXX$_\frac{1}{2}$ model is computed by inserting the values for $\mathcal{m}_i$ and $\mathcal{D}_i$ into Eqs.\,(\ref{eq:local_rho_matrix}) to (\ref{eq:spin_rho}).

The first observation is that while both the RISB and gRISB calculations share a dome-like structure in $I(\,\up\,:\,\dw\,)$, with the noninteracting and the strong-coupling limits giving a local free state, the Heisenberg model calculations provide a large value of $I(\,\up\,:\,\dw\,)$ at large-$U$, not far from that of the paramagnetic Mott-Hubbard insulator (see
Fig.\,\ref{fig:square_mit}). 
We can understand this difference in terms of the behavior of 
$\mathcal{m}_i$ and $\mathcal{D}_i$, the only ingredients to compute the local nonfreeness in our formulation. 
The double occupancy $\mathcal{D}_i$ is correctly found as small but finite in gRISB, as opposed to the zero value of the Heisenberg model.
On the other hand, $\mathcal{m}_i$ is substantially reduced in the Heisenberg model with respect to RISB and gRISB, as a result of nonlocal effects~\cite{Reger_squareXXX_Apr1988,Horsch_squareXXX_Jun1988,Hirsch_square_Jun1988,Gross_squareXXX_1989,Reger_2dXXX_1989,Sandvik_squareXXX_2002,Sandvik_honeyXXX_2006}, absent in (g)RISB. This is remarkably reflected also in $I(\,\up\,:\,\dw\,)$, a quantity based on measurements defined on the local (natural) spin orbitals.
Thus, local correlations are clearly influenced by nonlocal quantum processes, as expected for any open subsystem embedded in a nondegenerate quantum ground state, as its local classical correlations must be understood as originating from (nonlocal) entanglement with its environment~\cite{Discord_RMP[qfromclassical],Artiaco_PRL2025}.


\begin{figure}
    \centering
    \includegraphics[width=.925\linewidth]{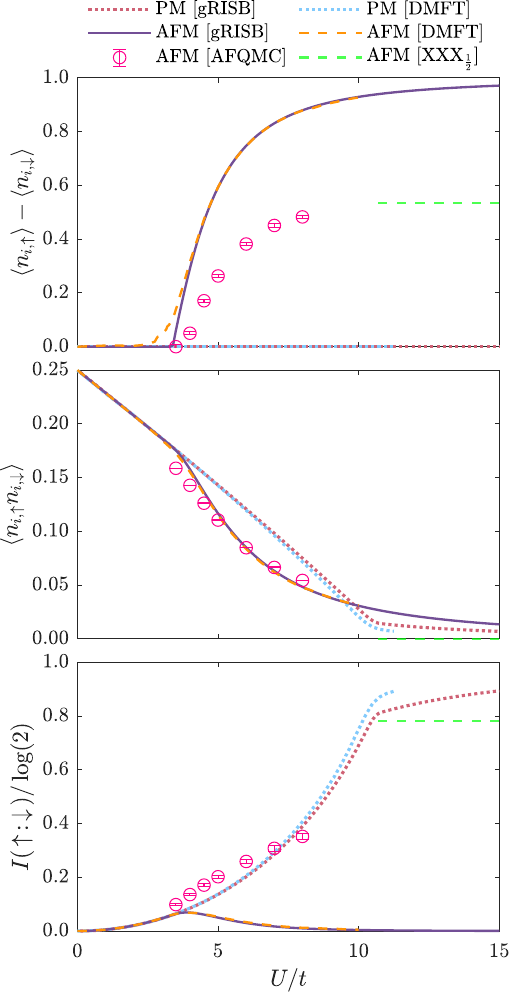}
    \caption{Local magnetization (top), local double occupancy (center) and local nonfreeness (bottom)
    on the honeycomb lattice, comparing the gRISB
    approximation ($N_\mathrm{ghost}=2$), the DMFT and AFQMC data. DMFT data are generated with the {EDIpack} 
    library \cite{DMFT/ED_Caffarel,DMFT/ED_Capone,EDIpack,EDIpack2,*EDIpack_code}.
    The AFQMC data are taken, 
    with permission, from \cite{DMFT/NRG_HoneyHubbard}. The error bars for the local mutual
    information are computed with linear propagation of the standard deviation.
    The dashed green line in the strong-coupling regime marks the numerically exact solution of the isotropic
    Heisenberg model on the honeycomb lattice, as reported in \cite{Sandvik_honeyXXX_2006}.}
    \label{fig:honey}
\end{figure}

We complete our analysis considering the honeycomb lattice, where the noninteracting system is a semimetal with Dirac cones. 
Hence, a finite critical value of $U$ is necessary to obtain an AFM state~\cite{SorellaTosatti92,CDMFT_HoneyHubbard,NoQSL_Honeycomb_Sorella,NoQSL_Honeycomb_Senechal,DCA_HoneyHubbard,DMFT/NRG_HoneyHubbard}. Also in this case, by inhibiting the Néel order, a paramagnetic
transition between a Dirac semimetal and a Mott-Hubbard insulator can be realized~\cite{Jafari2009,DMFT/ED_HoneyHubbard_SMIT}.
In Fig.\,\ref{fig:honey} we present results for both PM and AFM transitions, comparing solutions from the gRISB and DMFT approaches, the spin-$\frac{1}{2}$ isotropic Heisenberg model (XXX$_\frac{1}{2}$) on the honeycomb lattice (referring to numerically exact results reported in~\cite{Sandvik_honeyXXX_2006}) and, finally, sign-free auxiliary-field quantum 
Monte Carlo (AFQMC) data for the low-temperature regime of the Hubbard model 
on the honeycomb lattice, taken from \cite{DMFT/NRG_HoneyHubbard}.
We find that gRISB and DMFT results are substantially equivalent for both
transitions, except for a slight discrepancy in the large-$U$ degree of correlation,
inside the PM Mott insulator \footnote{Yet, the implied observation that gRISB
could capture less correlations than DMFT, with two ghost orbitals, appears to be in good
agreement with some conjectures on the relationship between the two methods \cite{Lanata_gRISB_vs_DMFT}.}.
On the other hand, while the AFQMC double occupancy in the large-$U$ 
regime is close to that of local embedding methods, the staggered magnetization is remarkably 
different and compatible with the Heisenberg prediction.
We can thus expect that $I(\,\up\,:\,\dw\,)$ would also asymptotically approach the 
very high degree of  correlation of the Heisenberg limit, which for the honeycomb lattice 
is of the order of magnitude of the paramagnetic Mott insulator 
(see the bottom panel in Fig.\,\ref{fig:honey}).

Our results show that quantum embedding methods based on a single-site do not fully
capture the substantial local classical correlations of the AFM ground state both on the
square and on the honeycomb lattice. The AFM local RDM turns out to be uncorrelated
like in a Hartree-Fock (Slater) magnetic state. This may appear in contrast with
nontrivial properties of the Mott antiferromagnet in
DMFT~\cite{Sangiovanni2006,Taranto2012,DMFT/NRG_HoneyHubbard}, which indeed
recovers the correct strong-coupling behavior of the ordering temperature, in
contrast with a simple Hartree-Fock treatment.
This behavior is however encoded in the nontrivial properties of the excited states,
in particular in the paramagnetic states found at high energy or temperature in DMFT,
while the ground state is essentially trivial and uncorrelated, at large $U$.

\clearpage
\newpage

\section*{Summary and outlook}
We provided an alternative way to evaluate the local nonfreeness in the framework of orbital quantum information theory. 
Remarkably, this allows to classify the electron-electron correlations in
classical and quantum contributions. 

For the Hubbard model (and closely related models where the orbital-resolved density and magnetization are conserved) this leads us to prove a remarkable and completely general result, namely that the local reduced density matrix describes only classical correlations between electrons, as it cannot contain electron-electron entanglement
or any kind of quantum correlation beyond entanglement.
The resulting measure of local classical correlations is shown to be uniquely defined and to vanish for any single-particle Hartree-Fock solution.

We apply this result to highlight some fundamental differences between local embedding descriptions of the paramagnetic and antiferromagnetic phases of the Hubbard model.
The paramagnetic Mott phase is characterized by a large degree of correlation which approaches the asymptotic maximal value in the large-$U$ limit. This effect is indeed well captured by quantum embedding methods that include only a local self-energy, such as DMFT and gRISB. 
On the other hand, the same embedding methods find a weakly correlated local density matrix in the AFM state, in sharp contrast with the large correlations predicted by numerically exact theories, in the large-$U$ regime. 
This implies that nonlocal self-energy effects (neglected in the local embedding) are crucial to increase the local classical correlations between electrons. Treating a single site as an open system, our results provide a concrete and nontrivial example of how local classical correlations can originate from the nonlocal correlations with its quantum environment (\ie the rest of the lattice) \cite{Discord_RMP[qfromclassical],Artiaco_PRL2025}. 

We highlight that our measure of local classical correlations depends only 
on fundamental and simple local quantities. 
As such, it is readily available in many theoretical methods including
nonlocal self-energies \cite{ClusterRMP2005,DiagramRoutes_RMP,DMFT/NRG_HoneyHubbard,Bippus2025,Bippus2026} 
that would allow to shed light
on the precise nonlocal origin of the classical correlations. The intraorbital
mutual information is also experimentally accessible in analog quantum simulators,
based on ultracold atoms \cite{ultracold_SU(N),Cocchi_ColdAtoms,Tusi2022NatPhys},
as well as digital quantum circuits 
\cite{Troyer2015_circuitHubbard,Campbell2022_circuitHubbard,Kan2025_circuitHubbard}.
The latter represent one of the main proposals for the exploitation of strongly
correlated electron systems on the path to achieve a practical quantum advantage with 
current hardware \cite{Poulin2021_advantageHubbard,Wataru2024_advantageHubbard}.
In this respect, the recent interest in alternative quantum many-body resources
\cite{QResources_RMP} 
has unveiled the \emph{nonstabilizerness} \cite{QMagic_as_resource,ManyBodyMagic_PRXQ} 
and the \emph{nongaussianity} (a resource-theory homologue of the nonfreeness \cite{LyuBu24_nonGaussianity,Sierant2025_FAF,Mele_NonfreenessBounds}) 
as crucial ingredients for generating high quantum complexity in fermionic systems \cite{Lami2025_GaussFermiMagic,Mele_NonfreenessLearning,Lami2025_nongauss_doping}.
Therefore the 
$I(\,\up\,:\,\dw\,)\equiv\mathfrak{N}(\rho_i)$ correspondence not only provides
an efficient scheme to estimate the degree of quantum advantage inherent to
strong correlations, but also paves the way towards a better understanding
of the intricate relationship between \emph{a priori} distinct many-body resources.

Finally, we highlight that realistic descriptions of strongly correlated \emph{materials}
often call for the inclusion of multiorbital interaction terms that go beyond orbital-resolved
conservation laws \cite{Kanamori,Rubtsov2012,Monserrat2020}. Our findings suggest that, besides
the improved accuracy, these additional terms may introduce local quantum correlations and entanglement,
modifying the physical picture of the underlying quantum states \emph{at a qualitative level}. 
A future, quantitative, assessment of these aspects is prospected to be very relevant for the multidisciplinary
efforts towards both an improved fundamental understanding of strong correlations in solids and
the development of materials-based quantum technologies.

\begin{acknowledgements}
G.B. acknowledges insightful discussions with C. Schilling, C. Mejuto-Zaera and S. Giuli. 
M. Raczkowski and F. Assaad kindly provided the lattice Monte Carlo data for the 
Hubbard model on the honeycomb lattice, originally published in \cite{DMFT/NRG_HoneyHubbard}.
All the RISB, gRISB and DMFT data are open available in \cite{DATASET_2506.18709}.
We acknowledge financial support from the National Recovery
and Resilience Plan PNRR MUR Project No.\,CN00000013-ICSC and by MUR via PRIN
2020 (Prot.\,2020JLZ52N-002) and PRIN 2022 (Prot.\,20228YCYY7) programs. A.A.
and M.C. acknowledge further financial support from the National Recovery and Resilience Plan
PNRR MUR Project No.\,PE0000023-NQSTI.
\end{acknowledgements}



\bibliography{main.bib}

\onecolumngrid
\vspace*{0.2cm}
\begin{center}
\textbf{APPENDIX \\[1ex] Local symmetries from global conservation of additive quantities}
\end{center}
\vspace{0cm}
\twocolumngrid

Here we prove the fundamental and general result that  the global conservation of density and magnetization reflect on the form of the local density matrix, thereby implying the classical character of the correlations.
Let us consider an \emph{additive} globally conserved quantity 
$\mathcal{O} = \sum_i \expval{\mathcal{O}_i}$, 
where $i$ spans a set of orbitals in the lattice.
For instance, we can always write the total density and the total magnetization in the Hubbard model 
as respectively $\mathcal{n} = \sum_i \expval{\mathcal{n}_i}$ 
and $\mathcal{m} = \sum_i \expval{\mathcal{m}_i}$, where 
$\expval{\mathcal{m}_i} = \expval{\mathcal{n}_{i,\up}}-\expval{\mathcal{n}_{i,\dw}}$.
Equivalently, we can consider the total spin-dependent densities
$\mathcal{n}_\sigma = \sum_i \expval{\mathcal{n}_{i,\sigma}}$. 

Similarly, we can write the total flavor-resolved density of a SU($N$) Hubbard model
\cite{ultracold_SU(N)} as a sum over all lattice sites
of the local flavor density
$\mathcal{n}_\nu = \sum_i \expval{\mathcal{n}_{i,\nu}}$,
or the total spin- and orbital-resolved density of a density-density 
Kanamori Hamiltonian \cite{Kanamori} as
$\mathcal{n} = \sum_i \expval{\mathcal{n}_{i,\nu,\sigma}}$.

All these quantities are described by \emph{local} operators $\mathcal{O}_i$, 
in the form:
\begin{equation}
   \expval{\mathcal{O}} = \expval{\mathcal{O}_1\otimes\mathbb{1}_2\otimes\dots}+
   \expval{\mathbb{1}_1\otimes \mathcal{O}_2\otimes\mathbb{1}_3\otimes\dots} + \dots \vspace{0.1cm}
   \label{eq:local_quantity}
\end{equation}
If $\expval{\mathcal{O}}$ is conserved on the lattice, the associated unitary group 
$U_\mathcal{O}=\mathrm{exp}(\mathrm{i}\alpha \mathcal{O})$ is factorized on local subspaces:
\begin{equation}
   U_\mathcal{O} = U_{\mathcal{O}_1}\otimes U_{\mathcal{O}_2} \otimes \dots =
   \mathrm{exp}(\mathrm{i}\alpha \mathcal{O}_1)\otimes\mathrm{exp}(\mathrm{i}\alpha \mathcal{O}_2)\dots
   \label{eq:local_unitary}
\end{equation}
It directly follows that the conservation of $\expval{\mathcal{O}}$, 
on the global state $\rho$, implies the conservation of $\expval{\mathcal{O}_i}$ 
on the local single-orbital reduced density matrix $\rho_i$, namely
$U_\mathcal{O}\,\rho\,U_\mathcal{O}^\dagger=\rho \Longrightarrow U_{\mathcal{O}_i}\,\rho_i\, U_{\mathcal{O}_i}^\dagger=\rho_i$. 
Hence, $\rho_i$ (or $\rho_{i,\nu}$ or
$\rho_{i,\nu,\sigma}$) must be diagonal over the local density and spin/flavors,
proving Eqs.\,(\ref{eq:rho_loc}) and (\ref{eq:zrho}) of the main text.

\end{document}